\pdfoutput=1
\documentclass[10pt,a4paper]{article}
\usepackage{times}
\usepackage[active]{srcltx}
\usepackage{graphics}
\usepackage{graphicx}
\newcommand{\si}{\mbox{\boldmath $\sigma$}}
\newcommand{\mb}{\mbox{\boldmath $\mu$}}
\begin{document}
\large
\date{ }

\begin{center}
{\Large Limits on short-range spin-dependent forces

 from spin relaxation of polarized $^{3}$He}

\vskip 0.7cm

Yu. N. Pokotilovski\footnote{e-mail: pokot@nf.jinr.ru}

\vskip 0.7cm
            Joint Institute for Nuclear Research\\
              141980 Dubna, Moscow region, Russia\\
\vskip 0.7cm

{\bf Abstract\\}

\begin{minipage}{130mm}

\vskip 0.7cm
 A new limit is presented on the axion-like monopole-dipole P,T-non-invariant
coupling in a range ($10^{-4} - 1$) cm.
 The gradient of spin-dependent nucleon-nucleon potential between $^{3}$He
nucleus and nucleons and electrons of the walls of a cell
containing polarized $^{3}$He gas should affect its spin
relaxation rate.
 The limit is obtained from the existing data on the relaxation rate of
spin-polarized $^{3}$He.

\end{minipage}
\end{center}
\vskip 0.3cm


\vskip 0.2cm


\vskip 0.6cm

 A number of proposals were published for the existence of new interactions
coupling mass to particle spin \cite {Lei,Hill,Fay,Dob}.
 On the other hand, there are theoretical indications that there may exist
light, scalar or pseudoscalar, weakly interacting bosons.
 Generally the masses and the coupling of these particles to nucleons,
leptons, and photons are not predicted by the proposed models.
 The most attractive solution of the strong CP problem is the existence of a
light pseudoscalar boson - the axion \cite{ax}.
 The axion may have a priori mass in a very large range, namely
$(10^{-12}<m_{a}<10^{6})$ eV.
 The main part of this mass range from both -- low and high mass boundaries --
was excluded as a result of numerous experiments and constraints from
astrophysical considerations \cite{win,PDG}.
 Astrophysical bounds are based on some assumptions concerning the axion and
photon fluxes produced in stellar plasma.
 These more recent constraints limit the axion mass to
$(10^{-5}<m_{a}<10^{-3})$ eV with small coupling constants to
quarks and photon \cite{win,PDG,rev}.
 Although these limits are more stringent than can be reached in laboratory
experiments, it is of interest to try to constrain the axion as much as
possible using laboratory means.
 The laboratory experiments performed or proposed so far are rather diverse
and employ a variety of detection techniques.
 The interpretation of laboratory experiments depend on less number of
assumptions than the constraints inferred from astrophysical and cosmological
observations and calculations.
 Axion is one of the best candidates for the cold dark matter of the Universe
\cite{dark}.

 Axions mediate a P- and T-reversal violating monopole-dipole interaction
potential between spin and matter (polarized and unpolarized nucleons)
\cite{Mood}:
%1
\begin{equation}
V({\bf r})=\si\cdot {\bf n}g_{s}g_{p}\kappa
\Bigl(\frac{1}{\lambda r}+\frac{1}{r^{2}}\Bigr)e^{-r/\lambda},
\end{equation}
where $g_{s}$ and $g_{p}$ are the dimensionless coupling constants of the
scalar and pseudoscalar vertices (unpolarized and polarized particles),
$\kappa=\hbar^{2}/(8\pi m_{n})$, $m_{n}$ is the nucleon mass at the polarized
vertex, $\si$ is the Pauli matrix related to spin of polarized nucleon,
$r$ is the distance between the nucleons,
$\lambda=\hbar/(m_{a}c)$ is the range of the force, $m_{a}$ - the axion mass,
and ${\bf n}={\bf r}$/r is the unitary vector directed from polarized nucleon
to unpolarized one.

 The potential between the layer of substance and the nucleon separated by the
distance $x$ from the surface is:
%2
\begin{equation}
V(x)=\mp 2\pi g_{s}g_{p}\kappa\lambda
N\,e^{-x/\lambda}(1-e^{-d/\lambda}),
\end{equation}
where $N$ is the nucleon density in the layer, $d$ is the layer's thickness.
 The "-" and "+" depend on the nucleon spin projection on x-axis (the surface
normal).

 Several laboratory searches provided constraints on axion-like
coupling in the macroscopic range $\lambda >0.1$ cm \cite{PDG}.

 The limit on this interaction in the $\lambda$--range $(10^{-4}-1)$ cm was
established in the Stern-Gerlach type experiment in which ultracold neutrons
(UCN) transmitted through a slit between a horizontal mirror and absorber
\cite{grav}.
 The obtained limit for the value $g_{s}g_{p}$ was $\sim 10^{-15}$ at
$\lambda=10^{-2}$ cm.
 This limit corresponds to the value of the monopole-dipole potential at the
surface of the mirror $\sim 10^{-3}$ neV, which is equivalent to the magnetic
field of $\sim 0.2$ G in the interaction $\mb{\bf H}$ of the neutron magnetic
moment with magnetic field.

 Sensitivity estimates for a future ultracold neutron Stern-Gerlach type
experiment were presented, which promise orders of magnitude improvements
in limiting the monopole-dipole interaction \cite{grav1}.

 There was also a proposal of the ultracold neutron magnetic resonance
frequency shift experiment for obtaining these constraints with better
precision \cite{Zimm}.

 It is shown here that constraints on this type of interaction may be obtained
from the existing experimental data on spin relaxation of polarized $^{3}$He.

 First, we consider a simple case of an infinite flat $^{3}$He cell.
 Two walls of this cell, in which polarized $^{3}$He gas is contained
between layers of thickness $d$, produce gradient of spin dependent potential:
%3
\begin{equation}
\frac{\partial V}{\partial x}=\pm 2\pi g_{s}g_{p}\kappa N
(1-e^{-d/\lambda})(e^{-x/\lambda}+e^{(x-L)/\lambda}),
\end{equation}
where $L$ is the distance between the walls (center of the cell is at $x=L/2$).

 The interaction energy ${\mb\bf H}$ of the particle magnetic moment in
a magnetic field is similar to the interaction energy $\si\bf H^{*}$
of the particle spin in the pseudo-magnetic monopole-dipole field $\bf H^{*}$
induced by nucleons in a substance.
 The action of the gradient of this field on the spin of polarized $^{3}$He is
equivalent to the action of the gradient of the magnetic field on the
magnetic moment.

 It is known that translational diffusion of polarized particles in the
chaotic magnetic fields affects significantly spin-relaxation, resulting in
the shortening of the spin-relaxation time.
 Physically it is explained by the fact that when a polarized particle
undergoes chaotic Brownian motion in the region of the magnetic field
gradients, it experiences randomly fluctuating magnetic fields.
 Spin-relaxation of atomic nuclei in gas depends strongly on these
fluctuations.
 The expression for the longitudinal spin-relaxation time $T_{1}$ in an
inhomogeneous magnetic field has been obtained in a number of works
(see \cite{Sche} and references therein).
 The rate of spin relaxation of $^{3}$He nuclei polarized along z-axis in the
gradient of magnetic field is
%4
\begin{equation}
\frac{1}{T_{1}^{grad}}=\frac{1}{3}
\frac{(\partial H_{x}/\partial x)^{2}+(\partial H_{y}/\partial y)^{2}}
{H_{z}^{2}}
<u^{2}>\frac{\tau_{c}}{1+(\omega_{0}\tau_{c})^{2}},
\end{equation}
where $<u^{2}>$ is the mean squared velocity of $^{3}$He atoms in a gas,
$\omega_{0}=2\mu H_{z}/\hbar$ is the magnetic resonance frequency in the
magnetic field applied along z-axis, $\tau_{c}$ is the time between
collisions of the $^{3}$He atoms in a gas.

 More general formula was derived in \cite{Cates} also valid at low magnetic
fields and low pressures.
 The critical parameter introduced in this work: $\omega_{0}R^{2}/D$, where
$R$ is the size of a cell, $D=u^{2}\tau_{c}/3$ is the $^{3}$He diffusion
coefficient in a gas, is $\sim 10^{6}$ in the experiments used to infer the
constraints on the axion-like coupling, and is very large at any reasonable
cell size and gas pressure.
 At large values of this parameter Eq. (4) is valid.

 When spin relaxation is caused by the gradient of spin-dependent potential
$V$, the rate of spin relaxation is
%5
\begin{equation}
\frac{1}{T_{1}^{grad}}= \frac{4}{3}
\frac{(\partial V_{x}/\partial x)^{2}+(\partial V_{y}/\partial y)^{2}}
{(\hbar\omega_{0})^{2}}<u^{2}>\frac{\tau_{c}}{1+(\omega_{0}\tau_{c})^{2}}.
\end{equation}

 For an infinite flat cell $V_{y}=0$.
 Averaging over the cell width gives (at $\omega_{0}\tau_{c}\ll 1$):
%6
\begin{equation}
\frac{1}{T_{1}^{grad}}=\frac{4}{3}\frac{(g_{s}g_{p}\kappa N)^{2}
<u^{2}>\tau_{c}}{(\hbar\omega_{0})^{2}} G_{inf},
\end{equation}
where
%7
\begin{equation}
G_{inf}=\frac{(2\pi)^{2}(1-e^{-d/\lambda})^{2}\lambda}{L}
\Bigl(1-e^{-2L/\lambda}+\frac{2L}{\lambda}e^{-L/\lambda}\Bigr).
\end{equation}

 It follows from Eqs. (3-7):
%8
\begin{equation}
g_{s}g_{p}=\Bigl(\frac{3}{4}\Bigr)^{1/2}
\frac{\hbar\omega_{0}}{\kappa N
(<u^{2}>\tau_{c}G_{inf}T^{grad}_{1})^{1/2}}.
\end{equation}

 For a finite cylindrical cell both $\partial V_{x}/\partial x$ and
$\partial V_{y}/\partial y$ components of pseudomagnetic potential are
essential.

 For a disc of radius $R$ and thickness $d$ with its axis along x-axis, the
potential at the point $\bf r$ is (Fig. 1):
%9
\begin{eqnarray}
V_{x}^{I disc}({\bf r})=g_{s}g_{p}\kappa N
\int_{0}^{2\pi}d\varphi\int_{0}^{R}\rho d\rho \int_{0}^{d}dt
\frac{-(x+t)}{q^{2}}\Bigl(\frac{1}{\lambda}+\frac{1}{q}\Bigr)
e^{-q/\lambda}, \nonumber\\
V_{y}^{I disc}({\bf r})=g_{s}g_{p}\kappa N
\int_{0}^{2\pi}d\varphi\int_{0}^{R}\rho d\rho \int_{0}^{d}dt
\frac{\beta}{q^{2}}\Bigl(\frac{1}{\lambda}+\frac{1}{q}\Bigr)
e^{-q/\lambda},
\end{eqnarray}
where $q=(r^{2}+\rho^{2}-2r\rho\cos{\varphi}+(x+t)^{2})^{1/2}$ is
the distance from the spin to the nucleus, $t$ is the distance
from the disc surface to the nucleus, $r=(y^{2}+z^{2})^{1/2}$ is the
projection of the radius-vector ${\bf r}$ of the spin on the $yz$-plane,
$\varphi_{1}$ is the angle between this projection and the
$z$-axis, $\rho$ is the projection of the radius-vector of the nucleus
on the $y,z$-plane, $\varphi$ is the angle between $r$ and $\rho$,
$\beta=\rho(\sin(\varphi_{1}+\varphi)+\cos(\varphi_{1}+\varphi))-
r(\sin{\varphi_{1}}+\cos{\varphi_{1}})$.

%========================================================
\begin{figure}
\begin{center}
%\resizebox{18cm}{9cm}{\includegraphics[0cm,0cm][24cm,12cm]{Fig1.eps}}
\resizebox{18cm}{12cm}{\includegraphics[width=\columnwidth]{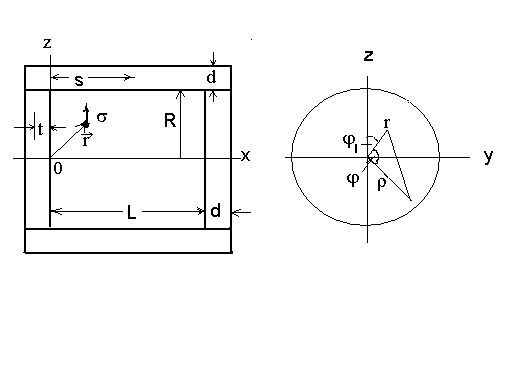}}
\end{center}
\caption{Geometry of a cylindrical cell used in calculations of $G_{cyl}$.}
\end{figure}
%============================================================

For the second disc of the cell $-(x+t)$ is replaced by $(L-x+t)$.

For a cylinder wall of internal radius $R$, length $L+2d$ and wall thickness
$d$
%10
\begin{eqnarray}
V_{x}^{cyl}({\bf r})=g_{s}g_{p}\kappa
\int_{0}^{2\pi}d\varphi\int_{R}^{R+d}\rho d\rho \int_{-d}^{L+d}ds
\frac{s-x}{q^{2}}\Bigl(\frac{1}{\lambda}+\frac{1}{q}\Bigr)
e^{-q/\lambda}, \nonumber\\
V_{y}^{cyl}({\bf r})=g_{s}g_{p}\kappa
\int_{0}^{2\pi}d\varphi\int_{R}^{R+d}\rho d\rho \int_{-d}^{L+d}ds
\frac{\beta}{q^{2}}\Bigl(\frac{1}{\lambda}+\frac{1}{q}\Bigr)
e^{-q/\lambda}.
\end{eqnarray}

 The derivatives of these potentials are the sums of contributions from all
walls of a cylindrical cell:
%11
\begin{eqnarray}
\frac{\partial V_{x}({\bf r})}{\partial x}=
\frac{\partial V_{x}^{I disc}({\bf r})}{\partial x}+
\frac{\partial V_{x}^{II disc}({\bf r})}{\partial x}+
\frac{\partial V_{x}^{cyl}({\bf r})}{\partial x}, \nonumber\\
\frac{\partial V_{y}({\bf r})}{\partial y}=
\frac{\partial V_{y}^{I disc}({\bf r})}{\partial y}+
\frac{\partial V_{y}^{II disc}({\bf r})}{\partial y}+
\frac{\partial V_{y}^{cyl}({\bf r})}{\partial y}.
\end{eqnarray}

 The sum of the squares of gradients averaged over the volume of the
cylindrical cell is
%12
\begin{equation}
\Bigl<\Bigl(\frac{\partial V_{x}}{\partial x}\Bigr)^{2}+
      \Bigl(\frac{\partial V_{y}}{\partial y}\Bigr)^{2}\Bigr>=
\frac{(g_{s}g_{p}\kappa N)^{2}}{V_{cell}}
\int_{0}^{2\pi}d\varphi_{1}\int_{0}^{R}r dr\int_{0}^{L}dx
\Bigl[\Bigl(\frac{\partial V_{x}({\bf r})}{\partial x}\Bigr)^{2}+
\Bigl(\frac{\partial V_{y}({\bf r})}{\partial y}\Bigr)^{2}\Bigr]=
(g_{s}g_{p}\kappa N)^{2}G_{cyl}.
\end{equation}

 The results of computation of $G_{cyl}$ when $R$=2.5 cm, $L$=5 cm and
$d$=0.2 cm are shown in Fig. 2 together with $G_{inf}$ at $L$=5 cm and
$d$=0.2 cm.
 It is seen that they coincide when $\lambda\ll R,L$.

%=================================================================
\begin{figure}
\begin{center}
%\resizebox{18cm}{9cm}{\includegraphics[0cm, 0cm][13cm,6cm]{Fig2.eps}}
\resizebox{18cm}{12cm}{\includegraphics[width=\columnwidth]{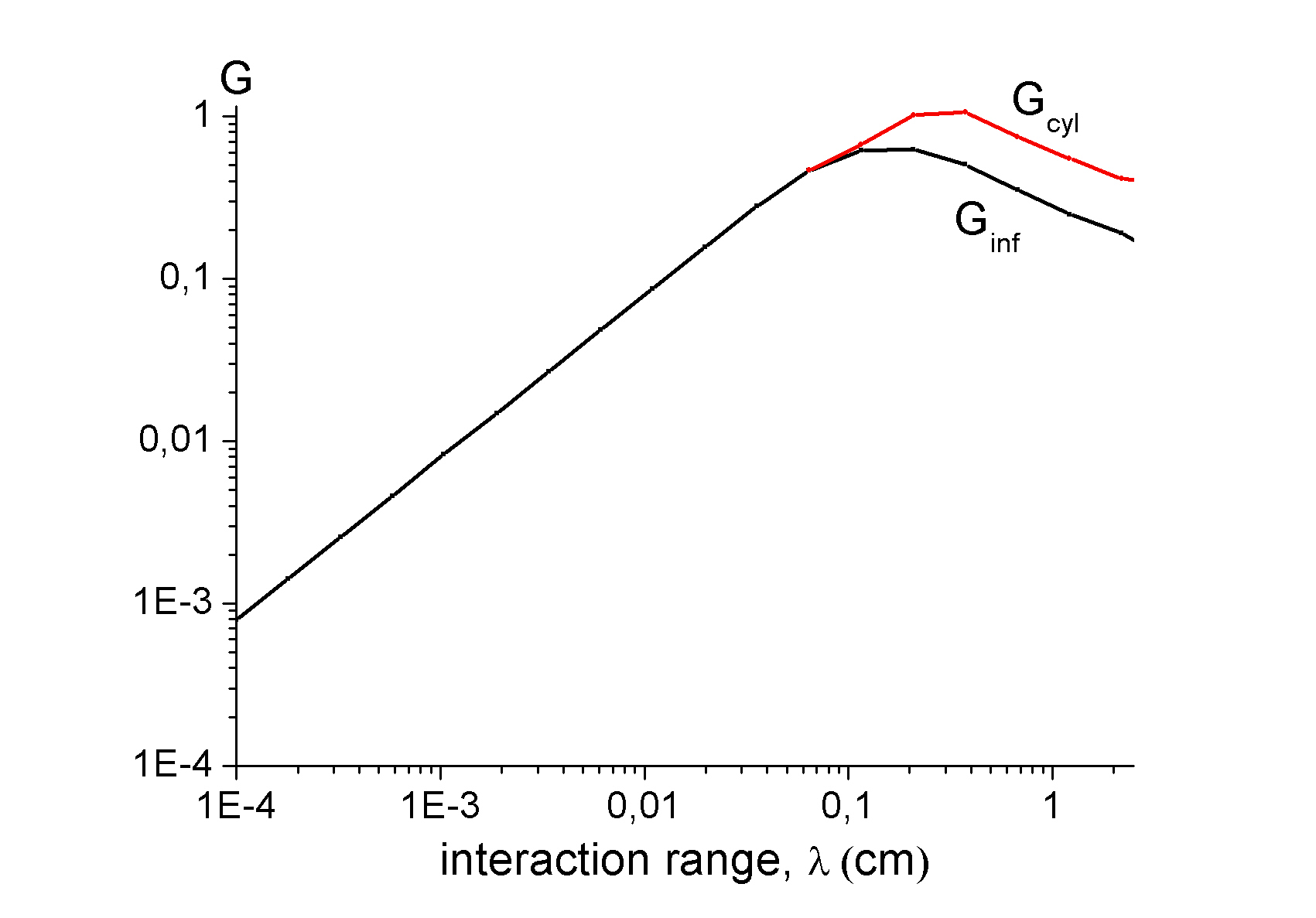}}
\end{center}
\caption{Calculated $G_{inf}$ (Eq. (7)) and $G_{cyl}$ (Eq. (12)).}
\end{figure}
%====================================================================

 Similarly to Eq. (6) for a cylindrical cell we have
%13
\begin{equation}
\frac{1}{T_{1}^{grad}}=\frac{4}{3}
\frac{(g_{s}g_{p}\kappa N)^{2}
<u^{2}>\tau_{c}}{(\hbar\omega_{0})^{2}} G_{cyl}
\end{equation}
and
%14
\begin{equation}
g_{s}g_{p}=\Bigl(\frac{3}{4}\Bigr)^{1/2}
\frac{\hbar\omega_{0}}{\kappa N
(<u^{2}>\tau_{c}G_{cyl}T^{grad}_{1})^{1/2}}.
\end{equation}

 Usually large variations of the $^{3}$He spin relaxation time are observed
among cells.
 For obtaining constraints for the monopole-dipole coupling we consider here
the results of recent measurements of the $^{3}$He spin relaxation
\cite{Rich,Gent,Parn,Gent1} in which the largest values of $T_{1}$ were
demonstrated.

 The cylindrical cell "1" \cite{Rich} has dimensions: diameter $4\times 5$ cm,
$^{3}$He pressure of 0.78 bar (corrected \cite{Gent1} compared to 0.85 bar
in the publication \cite{Rich}), and the spin relaxation time
$T_{1}^{exp}=840\pm 16$ hours.

 The cell "Diamond" \cite{Gent1} has spherical form, diameter 3 cm, $^{3}$He
pressure of 0.13 bar plus 0.9 bar of $^{4}$He, and the spin
relaxation time $T_{1}^{exp}=3000\pm 500$ hours.

 The cylindrical cell "j1" \cite{Parn} has dimensions: diameter $5\times 5$ cm,
pressure of 0.93 bar (corrected \cite{Mas} compared to 0.97 bar in the
publication \cite{Parn}), and spin relaxation time
$T_{1}^{exp}=663\pm 7$ hours.
% The magnetic field $H_{0}$ in these cells was 20 G, 20 G and
%10 G, respectively.

 The experimental spin relaxation time is determined by the contributions from
several random processes of time independent relaxation, the total
relaxation rate is the sum of rates for each process:
%15
\begin{equation}
\frac{1}{T_{1}^{exp}}=\frac{1}{T_{1}^{dip-dip}}+\frac{1}{T_{1}^{wall}}+
\frac{1}{T_{1}^{inhom}}+\frac{1}{T_{1}^{unknown}},
\end{equation}
where $T_{1}^{dip-dip}$ is the bulk dipole-dipole relaxation time,
$T_{1}^{wall}$ is due to the $^{3}$He spin relaxation on the walls of the
cell, $T_{1}^{inhom}$ is due to the magnetic field inhomogeneities,
$T_{1}^{unknown}$ may be determined by unknown factors.

 According to the calculations by Newbury et al. \cite{Newb} of the
magnetic-dipole interaction between nuclear spins in the $^{3}$He gas
$T_{1}^{dip-dip}=807/P$ hours, where $P$ is the $^{3}$He pressure in bar for
a temperature of 296 K.
 The precision of these calculations according to \cite{NewHap,New1} was
about 1\%.
 Possible contribution of any nonmagnetic dipole-dipole interaction between
$^{3}$He atoms is small compared to this uncertainty \cite{Ram,Vas}.

 We use here the published data for cylindrical cells of Refs.
\cite{Rich} and \cite{Parn}.
 The appropriate dipole-dipole relaxation rate has been subtracted from
these data.
 After this subtraction the remaining relaxation time is
$T_{1}^{rem}=4466\pm 245$ hours for Ref. \cite{Rich}, and
$T_{1}^{rem}=2810\pm 146$ hours for Ref. \cite{Parn}.
 In the calculation of uncertainties of $T_{1}^{rem}$ it was assumed that
the errors in the $^{3}$He pressure measurements performed by the neutron
transmission were about 5\% \cite{Gent1,NewHap}.
 As is seen, the remaining relaxation times for these cells are not
significantly different.

 These values of $T_{1}^{rem}$ were used for obtaining constraints on the
monopole-dipole interaction, the unknown value of wall relaxation rate being
attributed to the effect of the monopole-dipole potential.
 Magnetic field inhomogeneities in these measurements were very small but not
exactly known, their effect on spin relaxation was also attributed to the
effect of the monopole-dipole potential.

 Taking $<u^{2}>=3kT/m_{^{3}He}=2.35\times 10^{10}$ (cm/s)$^{2}$,
$\tau_{c}=3\times 10^{-10}$ s \cite{Newb},
$\omega_{0}=10^{5}$ s$^{-1}$, ($H_{z}=10\, G$,
the gyromagnetic ratio $\gamma_{^{3}He}=1.62$ kHz/G),
$N=1.5\times 10^{24}$ cm$^{-3}$,
the thickness of the glass walls of the cell d=0.2 cm,
the width of the cell L=5 cm \cite{Parn}, we get
%16
\begin{equation}
g_{s}g_{p}\approx \frac{8.4\times 10^{-16}}{(G_{cyl}T_{1}^{rem})^{1/2}}.
\end{equation}

 The obtained constraints are shown in Fig. 3 together with the constraints
known from other sources.

%=============================================================
\begin{figure}
\begin{center}
%\resizebox{18cm}{9cm}{\includegraphics[0cm, 0cm][13cm,6cm]{fig3.eps}}
\resizebox{18cm}{12cm}{\includegraphics[width=\columnwidth]{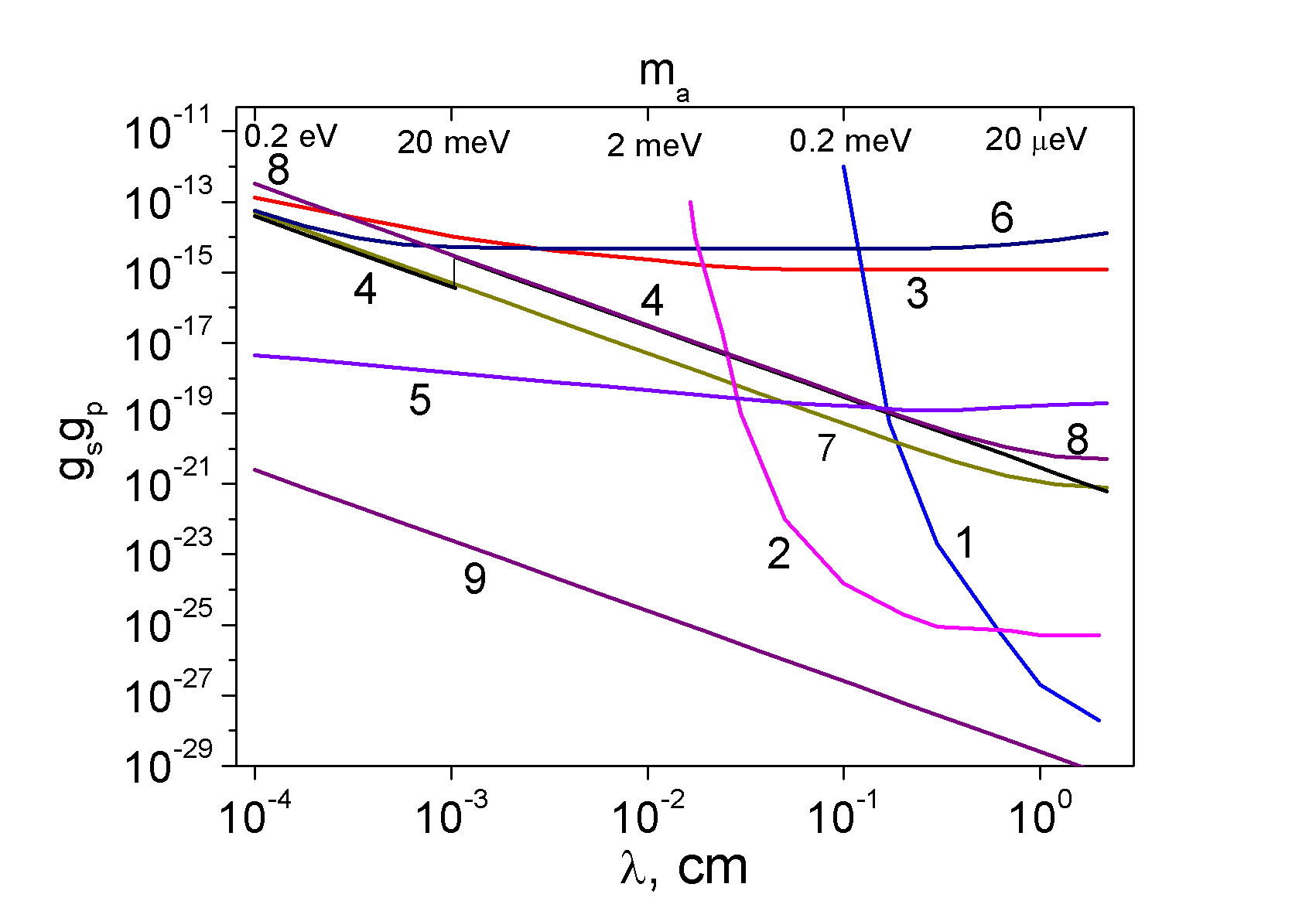}}
\end{center}
\caption{Constraints on the axion monopole-dipole coupling
strength $g_{s}g_{p}$ and effective range $\lambda$:
1 and 2 -  constraints for the value of coupling constant of nucleon and
electron $g_{s}^{n}g_{p}^{e}$ from Refs. \protect\cite{Ritt} and \protect\cite{Hamm},
respectively;
3 - from the UCN Stern-Gerlach experiment \protect\cite{grav};
4 - from the UCN depolarization probability according to \protect\cite{Ser};
5 - from spin relaxation of $^{3}He$, this work;
6, 7, and 8 - from the UCN depolarization probability \protect\cite{myUCN}, in
different assumptions regarding the experimental conditions of the UCN
depolarization measurement;
9 - from the product of separate constraints for $g_{s}$ from gravitational
experiments of the Seattle \protect\cite{Se04,Se07,Seconstr} and Stanford
\protect\cite{St03,St05,St08} groups, and astrophysical constraints on $g_{p}$
\protect\cite{Raf,PDG}.}
\end{figure}
%==================================================================

 The value of $T_{1}^{rem}$=2518 hours was used here -- two standard errors
less than the mean remaining longitudinal relaxation time from the
measurements \cite{Parn}.

 These $^{3}$He relaxation time data may be used to set
limits on the monopole-dipole coupling between nucleon spins of the
$^{3}$He nuclei and electrons of the walls of the cell.
 The density of electrons in the medium is approximately two times lower than
the density of nucleons, therefore the constraints are respectively two times
less strong.

 These constraints should be improved in dedicated experiments with polarized
$^{3}$He gas.
 First, if the wall relaxation could be further decreased,
better sensitivity would be obtained to additional sources of spin
relaxation in the $^{3}$He cells.
 At lower gas pressure the time between atom collisions $\tau_{c}$
is larger, which gives better sensitivity, but at the condition,
that the free path length between atom collisions in the gas cell
$u\tau_{c}\ll\lambda$.
 The geometry of a cell may be optimized for the chosen interaction range
$\lambda$.
 Generally, it would be good to use the narrowest possible cell,
for large $\lambda$ to place additional mass with the largest nucleon 
density in close vicinity to the walls of a cell.
 In the limit $d\gg\lambda\gg R,L$ we have $G_{inf}\rightarrow(4\pi)^{2}$.
 The sensitivity is increased also if to decrease the guiding
magnetic field $H_{z}$.

 Author is grateful to Yu. A. Plis for discussions, S. Masalovich for his
additional information on the experiment \cite{Parn}, W. M. Snow,
Changbo Fu and especially T. R. Gentile for their interest, the
information about the experiments \cite{Rich,Gent,Gent1} and
discussions, and to N. R. Newbury and W. Happer for their comments
on the dipole-dipole relaxation calculations and measurements.

\end{document}